\documentclass[manuscript]{aastex631}
\usepackage{amsmath}
\usepackage{soul}

\begin{document}

\title{The NANOGrav 12.5-year Data Set: Search for Gravitational Wave Memory}

\author[0000-0001-5134-3925]{Gabriella Agazie}
\affiliation{Center for Gravitation, Cosmology and Astrophysics, Department of Physics, University of Wisconsin-Milwaukee,\\ P.O. Box 413, Milwaukee, WI 53201, USA}
\author{Zaven Arzoumanian}
\affiliation{X-Ray Astrophysics Laboratory, NASA Goddard Space Flight Center, Code 662, Greenbelt, MD 20771, USA}
\author[0000-0003-2745-753X]{Paul T. Baker}
\affiliation{Department of Physics and Astronomy, Widener University, One University Place, Chester, PA 19013, USA}
\author[0000-0003-0909-5563]{Bence Bécsy}
\affiliation{Department of Physics, Oregon State University, Corvallis, OR 97331, USA}
\author[0000-0002-2183-1087]{Laura Blecha}
\affiliation{Physics Department, University of Florida, Gainesville, FL 32611, USA}
\author[0000-0003-4046-884X]{Harsha Blumer}
\affiliation{Department of Physics and Astronomy, West Virginia University, P.O. Box 6315, Morgantown, WV 26506, USA}
\affiliation{Center for Gravitational Waves and Cosmology, West Virginia University, Chestnut Ridge Research Building, Morgantown, WV 26505, USA}
\author[0000-0001-6341-7178]{Adam Brazier}
\affiliation{Cornell Center for Astrophysics and Planetary Science and Department of Astronomy, Cornell University, Ithaca, NY 14853, USA}
\affiliation{Cornell Center for Advanced Computing, Cornell University, Ithaca, NY 14853, USA}
\author[0000-0003-3053-6538]{Paul R. Brook}
\affiliation{Institute for Gravitational Wave Astronomy and School of Physics and Astronomy, University of Birmingham, Edgbaston, Birmingham B15 2TT, UK}
\author[0000-0003-4052-7838]{Sarah Burke-Spolaor}
\affiliation{Department of Physics and Astronomy, West Virginia University, P.O. Box 6315, Morgantown, WV 26506, USA}
\affiliation{Center for Gravitational Waves and Cosmology, West Virginia University, Chestnut Ridge Research Building, Morgantown, WV 26505, USA}
\author[0009-0008-3649-0618]{Rand Burnette}
\affiliation{Department of Physics, Oregon State University, Corvallis, OR 97331, USA}
\author[0009-0007-4346-8921]{Robin Case}
\affiliation{Department of Physics, Oregon State University, Corvallis, OR 97331, USA}
\author[0000-0002-5557-4007]{J. Andrew Casey-Clyde}
\affiliation{Department of Physics, University of Connecticut, 196 Auditorium Road, U-3046, Storrs, CT 06269-3046, USA}
\author[0000-0003-3579-2522]{Maria Charisi}
\affiliation{Department of Physics and Astronomy, Vanderbilt University, 2301 Vanderbilt Place, Nashville, TN 37235, USA}
\author[0000-0002-2878-1502]{Shami Chatterjee}
\affiliation{Cornell Center for Astrophysics and Planetary Science and Department of Astronomy, Cornell University, Ithaca, NY 14853, USA}
\author[0000-0001-7587-5483]{Tyler Cohen}
\affiliation{Department of Physics, New Mexico Institute of Mining and Technology, 801 Leroy Place, Socorro, NM 87801, USA}
\author[0000-0002-4049-1882]{James M. Cordes}
\affiliation{Cornell Center for Astrophysics and Planetary Science and Department of Astronomy, Cornell University, Ithaca, NY 14853, USA}
\author[0000-0002-7435-0869]{Neil J. Cornish}
\affiliation{Department of Physics, Montana State University, Bozeman, MT 59717, USA}
\author[0000-0002-2578-0360]{Fronefield Crawford}
\affiliation{Department of Physics and Astronomy, Franklin \& Marshall College, P.O. Box 3003, Lancaster, PA 17604, USA}
\author[0000-0002-6039-692X]{H. Thankful Cromartie}
\altaffiliation{NASA Hubble Fellowship: Einstein Postdoctoral Fellow}
\affiliation{Cornell Center for Astrophysics and Planetary Science and Department of Astronomy, Cornell University, Ithaca, NY 14853, USA}
\author[0000-0002-2185-1790]{Megan E. DeCesar}
\affiliation{George Mason University, resident at the Naval Research Laboratory, Washington, DC 20375, USA}
\author[0009-0009-3479-9897]{Dallas DeGan}
\affiliation{Department of Physics, Oregon State University, Corvallis, OR 97331, USA}
\author[0000-0002-6664-965X]{Paul B. Demorest}
\affiliation{National Radio Astronomy Observatory, 1003 Lopezville Rd., Socorro, NM 87801, USA}
\author[0000-0001-8885-6388]{Timothy Dolch}
\affiliation{Department of Physics, Hillsdale College, 33 E. College Street, Hillsdale, MI 49242, USA}
\affiliation{Eureka Scientific, 2452 Delmer Street, Suite 100, Oakland, CA 94602-3017, USA}
\author{Brendan Drachler}
\affiliation{School of Physics and Astronomy, Rochester Institute of Technology, Rochester, NY 14623, USA}
\affiliation{Laboratory for Multiwavelength Astrophysics, Rochester Institute of Technology, Rochester, NY 14623, USA}
\author{Justin A. Ellis}
\altaffiliation{Infinia ML, 202 Rigsbee Avenue, Durham NC, 27701}
\author[0000-0002-2223-1235]{Robert D. Ferdman}
\affiliation{School of Chemistry, University of East Anglia, Norwich, NR4 7TJ, United Kingdom}
\author[0000-0001-7828-7708]{Elizabeth C. Ferrara}
\affiliation{Department of Astronomy, University of Maryland, College Park, MD 20742}
\affiliation{Center for Research and Exploration in Space Science and Technology, NASA/GSFC, Greenbelt, MD 20771}
\affiliation{NASA Goddard Space Flight Center, Greenbelt, MD 20771, USA}
\author[0000-0001-5645-5336]{William Fiore}
\affiliation{Department of Physics and Astronomy, West Virginia University, P.O. Box 6315, Morgantown, WV 26506, USA}
\affiliation{Center for Gravitational Waves and Cosmology, West Virginia University, Chestnut Ridge Research Building, Morgantown, WV 26505, USA}
\author[0000-0001-8384-5049]{Emmanuel Fonseca}
\affiliation{Department of Physics and Astronomy, West Virginia University, P.O. Box 6315, Morgantown, WV 26506, USA}
\affiliation{Center for Gravitational Waves and Cosmology, West Virginia University, Chestnut Ridge Research Building, Morgantown, WV 26505, USA}
\author[0000-0001-7624-4616]{Gabriel E. Freedman}
\affiliation{Center for Gravitation, Cosmology and Astrophysics, Department of Physics, University of Wisconsin-Milwaukee,\\ P.O. Box 413, Milwaukee, WI 53201, USA}
\author[0000-0001-6166-9646]{Nate Garver-Daniels}
\affiliation{Department of Physics and Astronomy, West Virginia University, P.O. Box 6315, Morgantown, WV 26506, USA}
\affiliation{Center for Gravitational Waves and Cosmology, West Virginia University, Chestnut Ridge Research Building, Morgantown, WV 26505, USA}
\author[0000-0001-8158-683X]{Peter A. Gentile}
\affiliation{Department of Physics and Astronomy, West Virginia University, P.O. Box 6315, Morgantown, WV 26506, USA}
\affiliation{Center for Gravitational Waves and Cosmology, West Virginia University, Chestnut Ridge Research Building, Morgantown, WV 26505, USA}
\author[0000-0003-4090-9780]{Joseph Glaser}
\affiliation{Department of Physics and Astronomy, West Virginia University, P.O. Box 6315, Morgantown, WV 26506, USA}
\affiliation{Center for Gravitational Waves and Cosmology, West Virginia University, Chestnut Ridge Research Building, Morgantown, WV 26505, USA}
\author[0000-0003-1884-348X]{Deborah C. Good}
\affiliation{Department of Physics, University of Connecticut, 196 Auditorium Road, U-3046, Storrs, CT 06269-3046, USA}
\affiliation{Center for Computational Astrophysics, Flatiron Institute, 162 5th Avenue, New York, NY 10010, USA}
\author[0000-0002-1146-0198]{Kayhan Gültekin}
\affiliation{Department of Astronomy and Astrophysics, University of Michigan, Ann Arbor, MI 48109, USA}
\author[0000-0003-2742-3321]{Jeffrey S. Hazboun}
\affiliation{Department of Physics, Oregon State University, Corvallis, OR 97331, USA}
\author[0000-0003-1082-2342]{Ross J. Jennings}
\altaffiliation{NANOGrav Physics Frontiers Center Postdoctoral Fellow}
\affiliation{Department of Physics and Astronomy, West Virginia University, P.O. Box 6315, Morgantown, WV 26506, USA}
\affiliation{Center for Gravitational Waves and Cosmology, West Virginia University, Chestnut Ridge Research Building, Morgantown, WV 26505, USA}
\author[0000-0002-7445-8423]{Aaron D. Johnson}
\affiliation{Center for Gravitation, Cosmology and Astrophysics, Department of Physics, University of Wisconsin-Milwaukee,\\ P.O. Box 413, Milwaukee, WI 53201, USA}
\affiliation{Division of Physics, Mathematics, and Astronomy, California Institute of Technology, Pasadena, CA 91125, USA}
\author[0000-0001-6607-3710]{Megan L. Jones}
\affiliation{Center for Gravitation, Cosmology and Astrophysics, Department of Physics, University of Wisconsin-Milwaukee,\\ P.O. Box 413, Milwaukee, WI 53201, USA}
\author[0000-0002-3654-980X]{Andrew R. Kaiser}
\affiliation{Department of Physics and Astronomy, West Virginia University, P.O. Box 6315, Morgantown, WV 26506, USA}
\affiliation{Center for Gravitational Waves and Cosmology, West Virginia University, Chestnut Ridge Research Building, Morgantown, WV 26505, USA}
\author[0000-0001-6295-2881]{David L. Kaplan}
\affiliation{Center for Gravitation, Cosmology and Astrophysics, Department of Physics, University of Wisconsin-Milwaukee,\\ P.O. Box 413, Milwaukee, WI 53201, USA}
\author[0000-0002-6625-6450]{Luke Zoltan Kelley}
\affiliation{Department of Astronomy, University of California, Berkeley, 501 Campbell Hall \#3411, Berkeley, CA 94720, USA}
\author[0000-0003-0123-7600]{Joey S. Key}
\affiliation{University of Washington Bothell, 18115 Campus Way NE, Bothell, WA 98011, USA}
\author[0000-0002-9197-7604]{Nima Laal}
\affiliation{Department of Physics, Oregon State University, Corvallis, OR 97331, USA}
\author[0000-0003-0721-651X]{Michael T. Lam}
\affiliation{SETI Institute, 339 N Bernardo Ave Suite 200, Mountain View, CA 94043, USA}
\affiliation{School of Physics and Astronomy, Rochester Institute of Technology, Rochester, NY 14623, USA}
\affiliation{Laboratory for Multiwavelength Astrophysics, Rochester Institute of Technology, Rochester, NY 14623, USA}
\author[0000-0003-1096-4156]{William G. Lamb}
\affiliation{Department of Physics and Astronomy, Vanderbilt University, 2301 Vanderbilt Place, Nashville, TN 37235, USA}
\author{T. Joseph W. Lazio}
\affiliation{Jet Propulsion Laboratory, California Institute of Technology, 4800 Oak Grove Drive, Pasadena, CA 91109, USA}
\author[0000-0003-0771-6581]{Natalia Lewandowska}
\affiliation{Department of Physics, State University of New York at Oswego, Oswego, NY, 13126, USA}
\author[0000-0001-5766-4287]{Tingting Liu}
\affiliation{Department of Physics and Astronomy, West Virginia University, P.O. Box 6315, Morgantown, WV 26506, USA}
\affiliation{Center for Gravitational Waves and Cosmology, West Virginia University, Chestnut Ridge Research Building, Morgantown, WV 26505, USA}
\author[0000-0003-1301-966X]{Duncan R. Lorimer}
\affiliation{Department of Physics and Astronomy, West Virginia University, P.O. Box 6315, Morgantown, WV 26506, USA}
\affiliation{Center for Gravitational Waves and Cosmology, West Virginia University, Chestnut Ridge Research Building, Morgantown, WV 26505, USA}
\author[0000-0001-5373-5914]{Jing Luo}
\altaffiliation{Deceased}
\affiliation{Department of Astronomy \& Astrophysics, University of Toronto, 50 Saint George Street, Toronto, ON M5S 3H4, Canada}
\author[0000-0001-5229-7430]{Ryan S. Lynch}
\affiliation{Green Bank Observatory, P.O. Box 2, Green Bank, WV 24944, USA}
\author[0000-0002-4430-102X]{Chung-Pei Ma}
\affiliation{Department of Astronomy, University of California, Berkeley, 501 Campbell Hall \#3411, Berkeley, CA 94720, USA}
\affiliation{Department of Physics, University of California, Berkeley, CA 94720, USA}
\author[0000-0003-2285-0404]{Dustin R. Madison}
\affiliation{Department of Physics, University of the Pacific, 3601 Pacific Avenue, Stockton, CA 95211, USA}
\author[0000-0001-5481-7559]{Alexander McEwen}
\affiliation{Center for Gravitation, Cosmology and Astrophysics, Department of Physics, University of Wisconsin-Milwaukee,\\ P.O. Box 413, Milwaukee, WI 53201, USA}
\author[0000-0002-2885-8485]{James W. McKee}
\affiliation{E.A. Milne Centre for Astrophysics, University of Hull, Cottingham Road, Kingston-upon-Hull, HU6 7RX, UK}
\affiliation{Centre of Excellence for Data Science, Artificial Intelligence and Modelling (DAIM), University of Hull, Cottingham Road, Kingston-upon-Hull, HU6 7RX, UK}
\author[0000-0001-7697-7422]{Maura A. McLaughlin}
\affiliation{Department of Physics and Astronomy, West Virginia University, P.O. Box 6315, Morgantown, WV 26506, USA}
\affiliation{Center for Gravitational Waves and Cosmology, West Virginia University, Chestnut Ridge Research Building, Morgantown, WV 26505, USA}
\author[0000-0002-2689-0190]{Patrick M. Meyers}
\affiliation{Division of Physics, Mathematics, and Astronomy, California Institute of Technology, Pasadena, CA 91125, USA}
\author[0000-0002-4307-1322]{Chiara M. F. Mingarelli}
\affiliation{Center for Computational Astrophysics, Flatiron Institute, 162 5th Avenue, New York, NY 10010, USA}
\affiliation{Department of Physics, University of Connecticut, 196 Auditorium Road, U-3046, Storrs, CT 06269-3046, USA}
\affiliation{Department of Physics, Yale University, New Haven, CT 06520, USA}
\author[0000-0003-2898-5844]{Andrea Mitridate}
\affiliation{Deutsches Elektronen-Synchrotron DESY, Notkestr. 85, 22607 Hamburg, Germany}
\author[0000-0002-3616-5160]{Cherry Ng}
\affiliation{Dunlap Institute for Astronomy and Astrophysics, University of Toronto, 50 St. George St., Toronto, ON M5S 3H4, Canada}
\author[0000-0002-6709-2566]{David J. Nice}
\affiliation{Department of Physics, Lafayette College, Easton, PA 18042, USA}
\author[0000-0002-4941-5333]{Stella Koch Ocker}
\affiliation{Cornell Center for Astrophysics and Planetary Science and Department of Astronomy, Cornell University, Ithaca, NY 14853, USA}
\author[0000-0002-2027-3714]{Ken D. Olum}
\affiliation{Institute of Cosmology, Department of Physics and Astronomy, Tufts University, Medford, MA 02155, USA}
\author[0000-0001-5465-2889]{Timothy T. Pennucci}
\affiliation{Institute of Physics and Astronomy, E\"{o}tv\"{o}s Lor\'{a}nd University, P\'{a}zm\'{a}ny P. s. 1/A, 1117 Budapest, Hungary}
\author[0000-0002-8826-1285]{Nihan S. Pol}
\affiliation{Department of Physics and Astronomy, Vanderbilt University, 2301 Vanderbilt Place, Nashville, TN 37235, USA}
\author[0000-0001-5799-9714]{Scott M. Ransom}
\affiliation{National Radio Astronomy Observatory, 520 Edgemont Road, Charlottesville, VA 22903, USA}
\author[0000-0002-5297-5278]{Paul S. Ray}
\affiliation{Space Science Division, Naval Research Laboratory, Washington, DC 20375-5352, USA}
\author[0000-0003-4915-3246]{Joseph D. Romano}
\affiliation{Department of Physics, Texas Tech University, Box 41051, Lubbock, TX 79409, USA}
\author[0009-0006-5476-3603]{Shashwat C. Sardesai}
\affiliation{Center for Gravitation, Cosmology and Astrophysics, Department of Physics, University of Wisconsin-Milwaukee,\\ P.O. Box 413, Milwaukee, WI 53201, USA}
\author[0000-0003-2807-6472]{Kai Schmitz}
\affiliation{Institute for Theoretical Physics, University of Münster, 48149 Münster, Germany}
\author[0000-0002-7778-2990]{Xavier Siemens}
\affiliation{Department of Physics, Oregon State University, Corvallis, OR 97331, USA}
\affiliation{Center for Gravitation, Cosmology and Astrophysics, Department of Physics, University of Wisconsin-Milwaukee,\\ P.O. Box 413, Milwaukee, WI 53201, USA}
\author[0000-0003-1407-6607]{Joseph Simon}
\altaffiliation{NSF Astronomy and Astrophysics Postdoctoral Fellow}
\affiliation{Department of Astrophysical and Planetary Sciences, University of Colorado, Boulder, CO 80309, USA}
\author[0000-0002-1530-9778]{Magdalena S. Siwek}
\affiliation{Center for Astrophysics, Harvard University, 60 Garden St, Cambridge, MA 02138}
\author[0000-0002-5176-2924]{Sophia V. Sosa Fiscella}
\affiliation{School of Physics and Astronomy, Rochester Institute of Technology, Rochester, NY 14623, USA}
\affiliation{Laboratory for Multiwavelength Astrophysics, Rochester Institute of Technology, Rochester, NY 14623, USA}
\author[0000-0002-6730-3298]{Renée Spiewak}
\affiliation{Jodrell Bank Centre for Astrophysics, University of Manchester, Manchester, M13 9PL, United Kingdom}
\author[0000-0001-9784-8670]{Ingrid H. Stairs}
\affiliation{Department of Physics and Astronomy, University of British Columbia, 6224 Agricultural Road, Vancouver, BC V6T 1Z1, Canada}
\author[0000-0002-1797-3277]{Daniel R. Stinebring}
\affiliation{Department of Physics and Astronomy, Oberlin College, Oberlin, OH 44074, USA}
\author[0000-0002-7261-594X]{Kevin Stovall}
\affiliation{National Radio Astronomy Observatory, 1003 Lopezville Rd., Socorro, NM 87801, USA}
\author[0000-0002-7933-493X]{Jerry P. Sun}
\affiliation{Department of Physics, Oregon State University, Corvallis, OR 97331, USA}
\author[0000-0002-1075-3837]{Joseph K. Swiggum}
\altaffiliation{NANOGrav Physics Frontiers Center Postdoctoral Fellow}
\affiliation{Department of Physics, Lafayette College, Easton, PA 18042, USA}
\author[0000-0001-9118-5589]{Jacob Taylor}
\affiliation{Department of Physics, Oregon State University, Corvallis, OR 97331, USA}
\author[0000-0003-0264-1453]{Stephen R. Taylor}
\affiliation{Department of Physics and Astronomy, Vanderbilt University, 2301 Vanderbilt Place, Nashville, TN 37235, USA}
\author[0000-0002-2451-7288]{Jacob E. Turner}
\affiliation{Department of Physics and Astronomy, West Virginia University, P.O. Box 6315, Morgantown, WV 26506, USA}
\affiliation{Center for Gravitational Waves and Cosmology, West Virginia University, Chestnut Ridge Research Building, Morgantown, WV 26505, USA}
\author[0000-0001-8800-0192]{Caner Unal}
\affiliation{Department of Physics, Ben-Gurion University of the Negev, Be'er Sheva 84105, Israel}
\affiliation{Feza Gursey Institute, Bogazici University, Kandilli, 34684, Istanbul, Turkey}
\author[0000-0002-4162-0033]{Michele Vallisneri}
\affiliation{Jet Propulsion Laboratory, California Institute of Technology, 4800 Oak Grove Drive, Pasadena, CA 91109, USA}
\affiliation{Division of Physics, Mathematics, and Astronomy, California Institute of Technology, Pasadena, CA 91125, USA}
\author[0000-0003-4700-9072]{Sarah J. Vigeland}
\affiliation{Center for Gravitation, Cosmology and Astrophysics, Department of Physics, University of Wisconsin-Milwaukee,\\ P.O. Box 413, Milwaukee, WI 53201, USA}
\author[0000-0001-9678-0299]{Haley M. Wahl}
\affiliation{Department of Physics and Astronomy, West Virginia University, P.O. Box 6315, Morgantown, WV 26506, USA}
\affiliation{Center for Gravitational Waves and Cosmology, West Virginia University, Chestnut Ridge Research Building, Morgantown, WV 26505, USA}
\author[0000-0002-6020-9274]{Caitlin A. Witt}
\affiliation{Center for Interdisciplinary Exploration and Research in Astrophysics (CIERA), Northwestern University, Evanston, IL 60208}
\affiliation{Adler Planetarium, 1300 S. DuSable Lake Shore Dr., Chicago, IL 60605, USA}
\author[0000-0002-0883-0688]{Olivia Young}
\affiliation{School of Physics and Astronomy, Rochester Institute of Technology, Rochester, NY 14623, USA}
\affiliation{Laboratory for Multiwavelength Astrophysics, Rochester Institute of Technology, Rochester, NY 14623, USA}

\collaboration{1000}{The NANOGrav Collaboration}
\noaffiliation
\correspondingauthor{Jerry Sun}
\email{jerry.sun@nanograv.org}

\begin{abstract}

We present the results of a Bayesian search for gravitational wave (GW) memory in the NANOGrav 12.5-yr data set. We find no convincing evidence for any gravitational wave memory signals in this data set (Bayes factor = 2.8). As such, we go on to place upper limits on the strain amplitude of GW memory events as a function of sky location and event epoch. These upper limits are computed using a signal model that assumes the existence of a common, spatially uncorrelated red noise in addition to a GW memory signal. The median strain upper limit as a function of sky position is approximately $3.3 \times 10^{-14}$. We also find that there are some differences in the upper limits as a function of sky position centered around PSR J0613$-$0200. This suggests that this pulsar has some excess noise which can be confounded with GW memory. Finally, the upper limits as a function of burst epoch continue to improve at later epochs. This improvement is attributable to the continued growth of the pulsar timing array.

\end{abstract}


\defcitealias{aggarwal_nanograv_2020}{NG11mem}
\defcitealias{arzoumanian_nanograv_2015}{NG5mem}
\defcitealias{arzoumanian_nanograv_2020}{NG12gwb}
\defcitealias{the_nanograv_collaboration_nanograv_2015}{NG9}
\defcitealias{arzoumanian_nanograv_2018}{NG11}
\defcitealias{alam_nanograv_2020}{NG12}

\section{Introduction} \label{sec:intro}

Any system which radiates gravitational waves (GWs) will also cause a permanent change in the spacetime metric. This effect, which was first derived in \citet{christodoulou_nonlinear_1991}, is called nonlinear gravitational wave memory. This effect was later understood to be the accumulation of memory from emitted GWs over the course of any GW-producing event \citep{wiseman_christodoulous_1991, thorne_gravitational-wave_1992, blanchet_hereditary_1992}. Much work has been done to estimate the size of the effects of nonlinear GW memory, and it has been shown that there is a reasonable chance that the GW memory effect is significant enough to be observed using modern GW detectors \citep{wiseman_christodoulous_1991, arun_25pn_2004,favata_nonlinear_2009, favata_post-newtonian_2009, favata_gravitational-wave_2010}. 

One such GW detector is a pulsar timing array (PTA). A PTA is a collection of millisecond pulsars (MSPs) which have extremely stable rotational periods \citep{lorimer_binary_2008}. Because of their stability, it is expected that by carefully observing times-of-arrival (TOAs) of radio pulses from these MSPs, it is possible to observe timing residuals induced by the passage of GWs \citep{sazhin_opportunities_1978, detweiler_pulsar_1979, hellings_upper_1983, Agazie_2023}. The combination of multiple MSPs into a PTA also offers boosted sensitivity when trying to detect GW signals that produce predictable correlations amongst multiple pulsars \citep{foster_constructing_1990, lommen_pulsar_2015}.  Currently, there are several PTA collaborations in operation, including the North American Nanohertz Observatory for Gravitational Waves \citep[NANOGrav,][]{mclaughlin_north_2013}, the European Pulsar Timing Array \citep[EPTA,][]{desvignes_high-precision_2016}, the Parkes Pulsar Timing Array \citep[PPTA,][]{manchester_parkes_2013}, and the Indian Pulsar Timing Array \citep[InPTA,][]{paul_indian_2019}. Together, these collaborations form the International Pulsar Timing Array \citep[IPTA,][]{verbiest_international_2016}.

In the absence of any exotic physics, PTAs are expected to first detect a gravitational wave background originating from an ensemble of supermassive black hole binary (SMBHB) systems, followed by continuous waves from particularly bright SMBHBs \citep{rosado_expected_2015}. Much work has already been done to characterize the GW background and place limits on continuous GWs \citep[e.g.,][]{arzoumanian_nanograv_2016, arzoumanian_nanograv_2018-1, arzoumanian_nanograv_2020, chen_common-red-signal_2021,antoniadis_international_2022, arzoumanian_nanograv_2023, falxa_searching_2023, Agazie_2023}. During the final inspiral and merger of a SMBHB system, the SMBHB strongly emits GWs which are outside the frequency band detectable by PTAs. However, the accumulated memory from these GWs may be significant enough to be detected by PTAs \citep{seto_search_2009, van_haasteren_gravitational-wave_2010,pshirkov_observing_2010,cordes_detecting_2012,madison_assessing_2014}. In addition to these signals, it is also possible to detect or constrain more exotic GW sources using PTA data. For example, cosmic strings are expected to emit strong bursts of GWs \citep{damour_gravitational_2000, siemens_gravitational_2006}. \citet{yonemaru_searching_2020} have placed limits on GW bursts from cosmic strings based on the second PPTA data release.

Several studies have already been done using PTA data to constrain GW memory. NANOGrav has published constraints on GW memory using their 5-year and 11-year data sets \citep[][hereafter \citetalias{arzoumanian_nanograv_2015} and \citetalias{aggarwal_nanograv_2020}, respectively]{arzoumanian_nanograv_2015, aggarwal_nanograv_2020}. The PPTA has also published constraints in \citet{wang_searching_2015}. \citet{madison_versatile_2016} have published the results of a search for GW memory from five galaxy clusters using PPTA data. 

Additionally, several studies have considered GW memory using ground- and space-based detectors like LIGO-Virgo-Kagra and LISA. \citet{lasky_detecting_2016} suggest a method to detect accumulated memory of many individual mergers, each of which is too weak to see by itself.  \citet{hubner_measuring_2020} found no evidence of GW memory using the ten binary black hole mergers detected by LIGO and Virgo during their first two observation runs.  \citet{boersma_forecasts_2020} forecasted that GW memory could be detected with total $\text{SNR} = 3$ after approximately five years of aLIGO operation at design sensitivity. \citet{favata_nonlinear_2009} estimated that memory from SMBHB mergers may be detectable ($\text{SNR} \sim 5$) out to $z \lesssim 2$ with LISA, and \citet{islo_prospects_2019} estimated that LISA may see between 1 and 10 memory events in its lifetime.

In this paper, we present our analysis of the NANOGrav 12.5-year data set \citep{alam_nanograv_2020} for GW memory. We find that there is no significant evidence for GW memory in the data set. The model including a memory signal is only very marginally favored, with a Bayes factor of $2.8$.  The posteriors from a full PTA  analysis show that there is a very weak hint for GW memory at three different epochs: MJDs 54000, 55400, and 57300. However, a more detailed analysis shows that these three features are spurious. Each event is only supported by one or two pulsars, and one even lies inside a data gap in which there are no TOAs. 

Thus, finding no GW memory events, we present upper limits constraining the amplitudes of any GW memory as functions of trial burst epoch and sky location. In addition, we use the constraints as a function of burst epoch to set constraints on rates of all astrophysical events which produce GW memory.

In \autoref{sec:data}, we will describe the NANOGrav 12.5-year data set. Then, in \autoref{sec:signal_model}, we will discuss how a GW memory wavefront affects TOAs from a PTA. Next, in \autoref{sec:methods}, we will summarize the mathematical techniques and software used in this search. Finally, we will discuss the results in \autoref{sec:results}.

\section{Data} \label{sec:data}
In this paper, we analyze the NANOGrav 12.5-year narrowband data set \citep{alam_nanograv_2020}. We will briefly summarize some key points about this data set, but more details may be found in \citet{alam_nanograv_2020} (hereafter  \citetalias{alam_nanograv_2020}).

This data set contains TOAs from observations of 47 pulsars made between July 2004 and June 2017. However, in this analysis, we used only the 45 pulsars with at least three years of data. These observations were performed using the Arecibo Observatory (AO) and Green Bank Telescope (GBT). All pulsars in the declination range $0^{\circ} < \delta < +39^{\circ}$ were observed at Arecibo, with the remaining pulsars observed at GBT. In addition, PSRs B1937+21 and J1713+0747, which lie within the aforementioned declination range, were also observed at GBT. Based on work by \citet{burt_optimizing_2011} and \citet{christy_optimization_2014}, six pulsars were also observed in a high-cadence program: PSRs J0030+0451, J1640+2224, J1713+0747, J1909$-$3744, J2043+1711, and J2317+1439. These six pulsars were observed weekly starting from 2013 at GBT and 2015 at AO. The remaining pulsars were observed monthly. 

Each pulsar was, where possible, observed using two different receivers at different frequency ranges to help understand and model out interstellar medium (ISM) and dispersion measure (DM) effects. At Arecibo, observations were performed with the 1.4 GHz receiver, and one of either the 430 MHz or 2.1 GHz receivers depending on the noise characteristics of the observed pulsar. At the GBT, monthly observations used the 1.4 GHz and 820 MHz receivers. Weekly observations, however, only used the 1.4 GHz receiver. The observations were initially recorded using the ASP/GASP backends at AO and GBT, respectively \citep{Demorest2007}. Later, between 2010 and 2012, these backends were replaced by the wideband backends PUPPI/GUPPI at AO and GBT, respectively \citep{duplain_data, ford_heterogeneous}. 

Each pulsar's timing model was fitted using \texttt{TEMPO}\footnote{https://tempo.sourceforge.net/}\citep{nice_tempo_2015} and checked for consistency using \texttt{TEMPO2}\footnote{https://bitbucket.org/psrsoft/tempo2
}\citep{hobbs_tempo2_2006}  and \texttt{PINT}\footnote{https://github.com/nanograv/PINT}\citep{luo_pint_2019}.

\section{Signal and Noise Model} \label{sec:signal_model}
In this section, we will discuss the effects of GW memory on TOA residuals in pulsar timing data and summarize all of the components of the signal model and noise model used in this search.

Qualitatively, a memory wavefront passing over a single pulsar will cause the observed rotational frequency of the pulsar to suddenly increase or decrease by a constant amount. A memory wavefront passing over the Earth will cause the observed rotational frequencies of \textit{all pulsars} in the PTA to suddenly increase or decrease by a constant amount. In either case, this sudden change to the observed rotational frequency will introduce timing residuals because of the difference between the pulsar's timing-model fitted rotational frequency and the observed rotational frequency. Because the difference between the expected frequency and observed frequency is a constant, this will cause residuals to accumulate linearly over the course of the observation. In the case of a memory wavefront passing over just one pulsar, we will only see residuals in the TOAs of that pulsar. If a wavefront passes over the earth, we will see the residuals begin accumulating in the TOAs of every single observed pulsar.

For a GW memory wavefront propagating in the direction $\hat{\mathbf{k}}$ with polarization angle $\psi$ passing over the line-of-sight to a pulsar at sky position $\mathbf{p}$, the residuals induced may be calculated 
 \citep{estabrook_response_1975, hellings_upper_1983}:

\begin{equation}\label{eqn:bwm_on_resids}
\delta t_{\text{mem}}(t) = B(\mathbf{\hat{k}}, \mathbf{\hat{p}}, \psi) \vspace{2mm} h_{\text{mem}}(t).
\end{equation}
The projection factor $B(\mathbf{\hat{k}}, \mathbf{\hat{p}}, \psi)$ accounts for the fact that the GW memory has a quadrupolar antenna pattern. The effect of the GW memory on the TOAs from a pulsar depends on where that pulsar lies inside the antenna pattern. For a pulsar at $\mathbf{p}$ and a wavefront propagating in the direction $\hat{\mathbf{k}}$ separated by an angle $\alpha$, we can write the projection factor as:

\begin{equation} \label{eqn:geometric_factor}
    B(\mathbf{\hat{k}}, \mathbf{\hat{p}}, \psi) = \frac{1}{2} \cos(2 \psi_{\hat{\mathbf{p}}})(1 - \cos{\alpha}),
\end{equation}
where $\alpha$ is the angle between $\hat{\mathbf{p}}$, and $\hat{\mathbf{k}}$, and $\psi_{\hat{\mathbf{p}}}$ is the angle between the principal polarization vector (defined by $\psi$) and the pulsar line-of-sight projected onto a plane perpendicular to $\hat{\mathbf{k}}$.

The second factor in \autoref{eqn:bwm_on_resids} carries the strength and time dependence of the burst. For a wavefront with a characteristic strain of $h_0$, we may write the time-dependence as

\begin{equation}\label{eqn:bwm_wavefront_strain}
h_{\text{mem}}(t) = h_0 \vspace{4mm} [(t - t_0) \Theta(t - t_0) - (t - t_i)\Theta(t - t_i)]
\end{equation}
where $t_0$ is the time at which the memory wavefront passes over the Earth, \hbox{$t_i = t_0 +(|\vec{\mathbf{p}_i}|/c) \hspace{1mm} [1 + \cos(\theta_i)]$} is the retarded time at which the same wavefront passed over the pulsar, and $\Theta$ is the Heaviside function. The left- and right-hand terms in \autoref{eqn:bwm_wavefront_strain} are called the ``Earth term'' and ``pulsar term'', respectively. In reality, because the distance to each pulsar in our PTA is on the order of thousands of light-years, and the observation baselines of ongoing PTA experiments is tens of years, we expect only one nonzero term in \autoref{eqn:bwm_wavefront_strain}. 

The characteristic strain of GW memory $h_0$ depends on the amount of energy radiated as GWs, $\Delta E_\mathrm{rad}$, and the orientation and distance of the source relative to the observer.  For a binary merger
\begin{equation}\label{eqn:characteristic_strain}
    h_{0} = \frac{\Delta E_{\text{rad}}}{24 r} \sin^2\iota \left(17 + \cos^2\iota\right),
\end{equation}
where $r$ is the comoving distance to the source, $\iota$ is the orbital inclination angle of the binary, and $\Delta E_\mathrm{rad}$ is a function of the individual masses and spins.  As our signal model includes only the memory portion, we are agnostic to the particulars of the signal's origin and parameterize our model with $h_0$ directly.

Because of the significance of the detection of a common spatially uncorrelated red noise (CURN) process in \citetalias{arzoumanian_nanograv_2020}, we also include a CURN with a fixed spectral index as part of our model. The CURN is modeled as a power law, with a power spectrum characterized by two hyperparameters $(A, \gamma)$ \citep{phinney_practical_2001}:

\begin{equation}\label{eqn:rn_plaw}
    P(f) = A^2\left(  \frac{f}{\text{yr}^{-1}} \right)^{-\gamma},
\end{equation}
where $f$ is the frequency of the spectral component, $A$ is the characteristic amplitude of the red-noise process at the reference frequency $\mathrm{yr}^{-1}$, and $\gamma$ is the spectral index of the process. A stochastic gravitational wave background generated by an ensemble of SMBHB is expected to have a spectral index of $\gamma_{\text{SMBHB}} = 13/3 \approx 4.33$. However, the maximum a posteriori value found for the spectral index of the CURN in \citetalias{arzoumanian_nanograv_2020} was approximately $\gamma_{\text{MAP}} = 5.5$. In this paper, we present two sets of results using both of these fixed spectral indices for the CURN.

In addition to these signals, we include Gaussian white noise and Gaussian red noise on a per-pulsar basis. The Gaussian red noise accounts for long-timescale changes in the pulsar's rotational frequency. Some examples of processes which can cause these changes include spin noise \citep{shannon_assessing_2010, lam_nanograv_2016-1}, stochastic variations in dispersion measure \citep{demorest_limits_2013, keith_measurement_2013, jones_nanograv_2017}, and mode changing \citep{lyne_switched_2010, miles_mode_2022}.

The Gaussian white noise is parameterized by three parameters known as EQUAD, EFAC, and ECORR. EQUAD and EFAC modify the measured TOA uncertainty: EQUAD adds additional white noise in quadrature, and EFAC multiplies the total TOA uncertainty after including EQUAD. ECORR describes white noise that is correlated between TOAs gathered in the same observation epoch but uncorrelated between different observations. This term nominally accounts for pulse jitter noise. For this analysis, the white noise parameters are fixed to their median values as determined by single pulsar noise analyses for the sake of computational efficiency.

\section{Methods} \label{sec:methods}
The techniques used in this search are documented in \citet{sun_implementation_2023}. As such, in this section, we will give only a brief overview of the techniques. The residuals in a single pulsar's TOAs may be written as the sum of multiple stochastic and deterministic processes:

\begin{equation}\label{eqn:residual_model}
    \mathbf{\delta t} = \mathbf{\delta t}_{\text{mem}} + M\boldsymbol\epsilon + F\mathbf{a} + F_{\text{gw}}\mathbf{a}_{\text{gw}} + \mathbf{n}.
\end{equation}
Above, $\mathbf{\delta t}$ are the residual timeseries for the pulsar. The term $\mathbf{\delta t}_{\text{mem}}$ are the residuals induced by GW memory; $M$ is the design matrix accounting for small errors in the linearized pulsar timing model $\boldsymbol\epsilon$; $F$ is the design matrix for a pulsar-intrinsic Gaussian red-noise process modeled as a Fourier series with coefficients $\mathbf{a}$; Similarly, $F_{\text{gw}}$ and $\mathbf{a}_{\text{gw}}$, are the design matrix and Fourier coefficients for the CURN; finally, $\mathbf{n}$ are the uncertainties in the TOAs from Gaussian white noise. 

Given estimations of the timing model parameters, GW memory signal, and Gaussian process parameters, we can construct residuals $\mathbf{r}$:

\begin{equation}\label{eqn:wn_resids}
    \mathbf{r} = \mathbf{\delta t} - \mathbf{\delta t}_{\text{mem}} - M\boldsymbol\epsilon - F\mathbf{a} -  F_{\text{gw}}\mathbf{a}_{\text{gw}}.
\end{equation}

Since the residuals $\mathbf{r}$ are expected to arise only from Gaussian white noise (having subtracted out all other effects), we can compute the likelihood of any set of model parameters as:

\begin{align}\label{eqn:likelihood}
\begin{split}
    L(\boldsymbol\epsilon, \mathbf{a}, \mathbf{a_{\text{gw}}}, \mathbf{\delta t}_{\text{mem}}) &= p( \mathbf{\delta t} | \boldsymbol\epsilon, \mathbf{a},  \mathbf{a_{\text{gw}}}, h_{\text{mem}}, t_0, \hat{\mathbf{k}},\hat{\mathbf{p}}, \psi) \\
     &= \frac{\exp{\left(-\frac{1}{2}\mathbf{r}^T N^{-1}\mathbf{r}\right)}}{\sqrt{2 \pi \det{N}}}.
\end{split}
\end{align}

It is also possible to analytically marginalize this likelihood over the timing model and red noise parameters. For a full description, see \citet{lentati_hyper-efficient_2013}, \citet{van_haasteren_new_2014}, and \citet{van_haasteren_low-rank_2015}. The final marginalized likelihood is:

\begin{equation}
p( \mathbf{\delta t} | h_{\text{mem}}, t_0, \hat{\mathbf{k}},\hat{\mathbf{p}}, \psi) = \frac{\exp{\left( -\frac{1}{2} \mathbf{q}^T C^{-1} \mathbf{q}\right)}}{\sqrt{2\pi \det{C}}}
\end{equation}
where we have the definitions:

\begin{equation}
\mathbf{q} = \delta \mathbf{t} - \delta \mathbf{t}_{\mathrm{mem}},
\end{equation}

\begin{equation}
C = N + T D T^T,
\end{equation}

\begin{equation}
  T =
  \left[ {\begin{array}{cc}
   M & F
  \end{array} } \right],
\end{equation}

\begin{equation}
  D =
  \left[ {\begin{array}{ccc}
   \infty & 0\\
    0 & \phi 
  \end{array} } \right],
\end{equation}
where $\infty$ above is a diagonal matrix of infinities (which we can understand as unconstrained priors on timing model parameters), and $\phi$ is a  covariance matrix for the individual red noise and CURN Fourier coefficients. Because we are using a CURN, these $\phi$ matrices are also diagonal, and simply contain the red noise power at each frequency bin given by \autoref{eqn:rn_plaw}.

Following the procedure of \citet{sun_implementation_2023}, which is based on methods in \S6 of \citetalias{arzoumanian_nanograv_2015}, we then compute the pulsar-term likelihoods (marginalized over intrinsic pulsar red noise and fixed spectral index CURN) on a grid of trial parameters $\{h_i, t_B\}$ where $h_i$ is the post-projection strain of the memory signal in the $i$-th pulsar, and $t_B$ is the burst epoch. The post-projection strain is given by the product of the projection factor (\autoref{eqn:geometric_factor}) and the intrinsic strain of the memory signal $h_0$. We can only use the post-projection strain since the residuals of one pulsar cannot break the degeneracy between the location of the signal's origin and the intrinsic strain. Additionally, \citet{sun_implementation_2023} do not include a CURN, but we choose to include this additional noise process because of the results of \citetalias{arzoumanian_nanograv_2020}, in which it was shown that there is significant evidence for a CURN in the NANOGrav 12.5-year data set \citepalias{alam_nanograv_2020}. These pulsar-term likelihood tables may be used to set upper limits on pulsar-term GW memory.

Then, we can combine the pulsar-term likelihoods to compute Earth-term likelihoods by making use of the factorizability of the signal model:

\begin{equation}\label{eqn:factorized_likelihood}
\begin{split}
    p(\mathbf{\delta t} | \hat{\mathbf{k}}, \psi, t_B, h_0) &= \prod_{i=1}^{N_{\text{psr}}}{p_{i}(\mathbf{\delta t} | \hat{\mathbf{k}}, \psi, t_B, h_{0}}) \\
    &= \prod_{i=1}^{N_{\text{psr}}}{p_i(\mathbf{\delta t}|h_{i}, t_B)},
\end{split}
\end{equation}
where above we have implicitly used \autoref{eqn:geometric_factor} to combine the burst parameters $\hat{\mathbf{k}}$, $\psi$, and $h_0$ into the post-projection, pulsar-term GW memory strain $h_i$. In this way, it becomes very computationally inexpensive to compute the red-noise-marginalized Earth-term likelihoods of any GW memory events on a full grid of trial parameters $\{ \log_{10}{h_0}, t_B , \theta_B, \phi_B, \psi_B\}$ where $h_0$ is the \textit{intrinsic} strain of the memory event, $t_B$ is the event epoch, $(\theta_B, \phi_B)$ are the polar and azimuthal angles of the sky location of the event source, and $\psi_B$ is the polarization angle of the memory wavefront. Once we have the likelihoods on a grid of trial parameters, we can simply numerically marginalize over any of the trial parameters to obtain marginalized likelihoods or posterior probability distributions. 

These signal models and this likelihood calculation are implemented in \texttt{enterprise}\footnote{https://github.com/nanograv/enterprise} \citep{ellis_justin_a_enterprise_2020} and \texttt{enterprise\_extensions}\footnote{https://github.com/nanograv/enterprise$\_$extensions} \citep{enterprise_extensions}

\section{Results}\label{sec:results}

\subsection{Earth-term Memory Search}
\begin{figure}    
    \centering
    \includegraphics[width=0.98\textwidth]{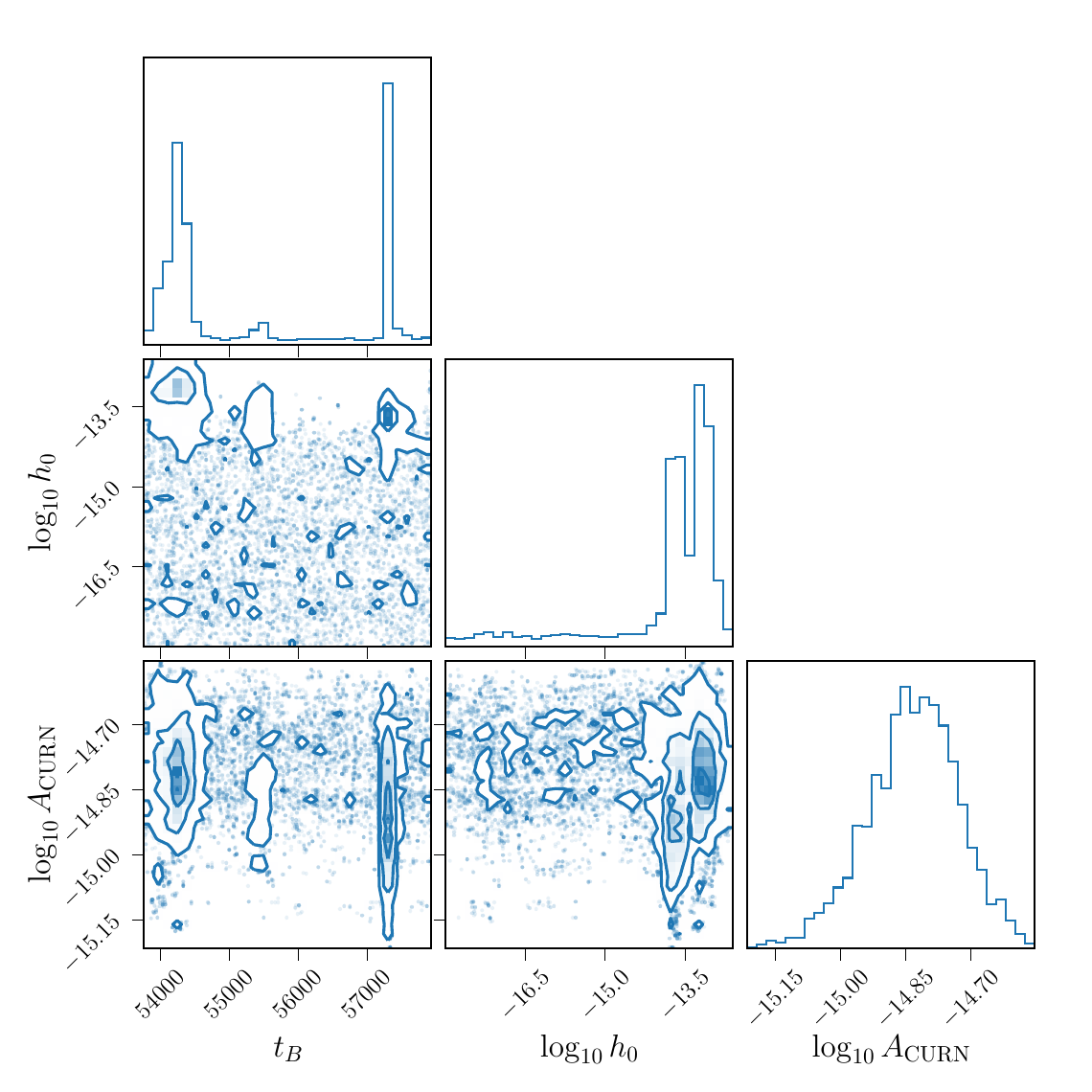}
    \caption{A corner plot showing 1D and 2D marginalized posteriors for three key model parameters: burst epoch $t_B$, burst strain amplitude $\log_{10} h_0$, and CURN amplitude $\log_{10} A_\mathrm{CURN}$.  The good localization $\log_{10} A_\mathrm{CURN}$ shows that the CURN is still detected in the presence of the memory model.  Furthermore, the tail of $\log_{10} h_0$ extends to very low amplitude, which indicates that $h_0 \sim 0$ is still supported by the model.}
    \label{fig:hypermodel_result}
\end{figure}

We began by performing a Earth-term Bayesian search for GW memory using MCMC sampling.  We compared two models: 1) a noise-only model and 2) a noise and GW memory model.  The noise-only model included intrinsic pulsar red noise, white noise, and a common red noise process.  The signal model included the same noise processes with an additional GW memory signal. The two models were simultaneously sampled using the product-space sampling method \citep{carlin_bayesian_1995,godsill_relationship_2001}, allowing us to determine the posterior probability for the memory signal and compute the Bayes factor for the signal model compared to noise only.
The resulting Bayes factor of $2.8$, shows the GW memory model is marginally favored over noise only. However, this Bayes factor is too small to be considered a detection.
The posterior probability distributions for the memory signal and global spatially uncorrelated red-noise process are shown in \autoref{fig:hypermodel_result}.  Based on the the posterior probability of the burst epoch, we can identify three ``hot spots'' near MJDs 54000, 55400 and 57300.

The features near MJD 54000 and 55400 were both present in the analysis of \citetalias{aggarwal_nanograv_2020}, where they were the most significant GW memory false alarm events in \citetalias{arzoumanian_nanograv_2018} and \citetalias{the_nanograv_collaboration_nanograv_2015}, respectively.
The feature near MJD 54000 lies near the start of our observations and at a time where there were large data gaps for several pulsars.  At early times in our dataset there were fewer pulsars being observed, and the observations were less regular.  The sparsity of data makes it harder to constrain any signal in these times.  Events that occur early in the data are also more degenerate with the quadratic pulsar timing model fit to the pulsar rotational frequency and frequency derivative.  This means that the signal model can be consistent with a high amplitude memory event that is effectively removed by the marginalization of the timing model.

Using a dropout analysis, we identified three pulsars in particular which supported each of the three aforementioned features: PSRs J0030+0451, J1744$-$1134 and J2043+1711. We performed another Bayesian search using a free spectral noise model, which treats the power in each frequency bin in the power spectral density as an independent parameter, rather than requiring a power-law red noise power spectral density for these three pulsars. We found that using a more flexible noise model in these pulsars completely removes the features at MJD 55400 and MJD 57300. Because each of these features has no support from any other pulsars, we conclude that these events are related to noise in individual pulsars and not actual GW memory events.
The analysis using the free spectral noise model in three pulsars results in a Bayes factor for GW memory of $1.3$. In general, more complex noise models like those used in \citet[][in prep]{adv_noise} should help prevent noise features from contaminating searches for GW memory in the future.

\subsection{Pulsar-term Upper Limits}

Because we make no detection, we report upper limits on GW memory strain amplitude. \label{sec:psrUL}
\begin{figure}   
    \centering
    \includegraphics[width=0.47\textwidth]{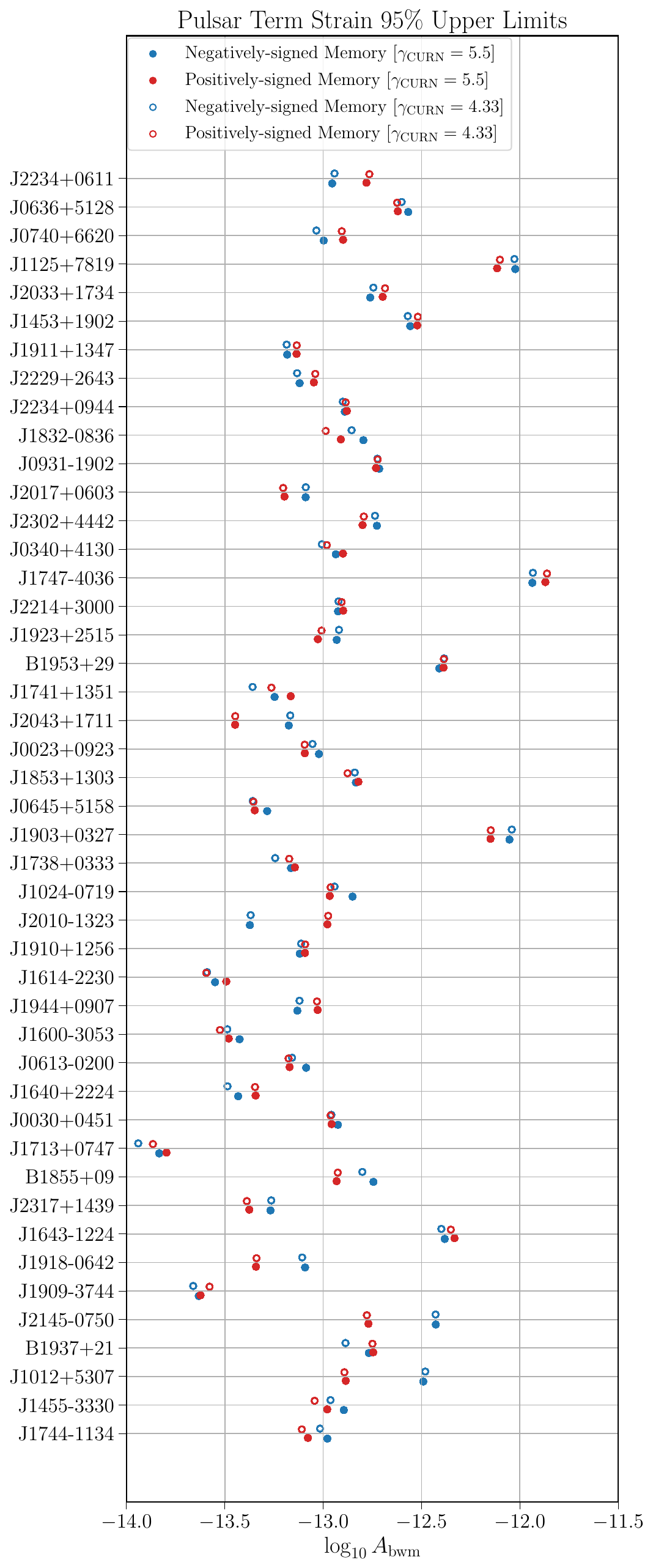}
    \caption{A plot of the pulsar-term upper limits on memory strain amplitude. The pulsars are listed in order of shortest to longest timing baseline. To find these upper limits, we compute amplitude posteriors from the pulsar-term lookup tables marginalized over the burst epoch, pulsar intrinsic red noise and a fixed spectral index common uncorrelated red noise (CURN) process. Overall, we do not find much difference in pulsar-term upper limits when comparing the results using a fixed CURN spectral index of $\gamma_{\text{SMBHB}} = 4.33$ and $\gamma_{\text{MAP}} = 5.5$.}
    \label{fig:psrterm_ULs_combined}
\end{figure}

\autoref{fig:psrterm_ULs_combined} shows the pulsar-term upper limits on GW memory using both $\gamma_{\text{MAP}}$ and $\gamma_{\text{SMBHB}}$. Because the pulsar-term upper limits are computed one pulsar at a time, we lose all information relating to the sky location of the signal. This amplitude upper limit is a constraint on the product of $B(\hat{\mathbf{k}}, \hat{\mathbf{p}}, \psi)$ and $h_0$, since these two terms are fully degenerate in the pulsar-term search. In other words, it is impossible to differentiate between a weak memory event or one that originated in the sky such that the antenna pattern is weak at the pulsar's sky location. We see that the choice of spectral index does not affect the pulsar-term upper limits very much in most cases. Some pulsars (e.g., PSRs B1937+21, J0613$-$0200, J0645+5158, J1713+0747) show small, but significant differences.

\subsection{Earth-term Upper Limits}
\label{sec:earthUL}
\begin{figure}
    \gridline{\fig{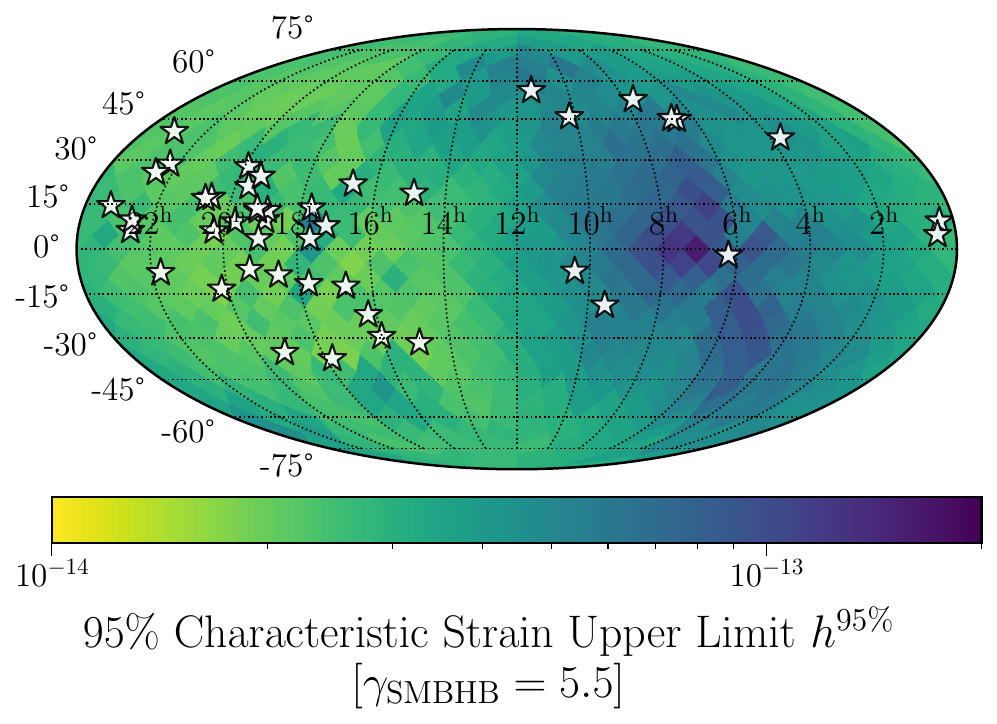}{0.48\textwidth}{}
              \fig{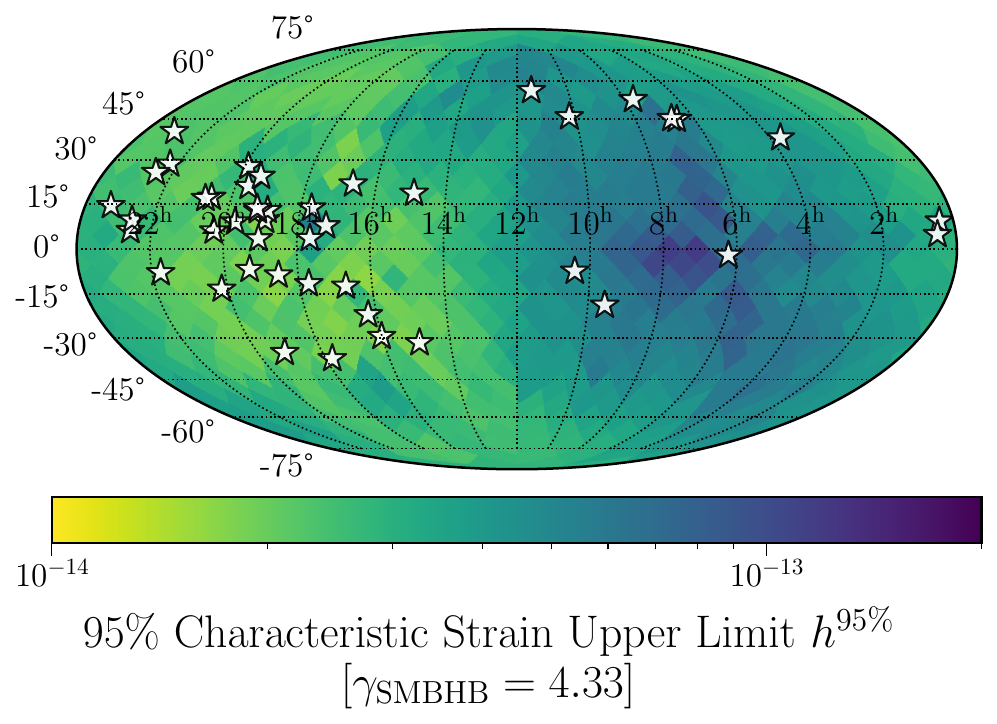}{0.48\textwidth}{}}
    \gridline{\fig{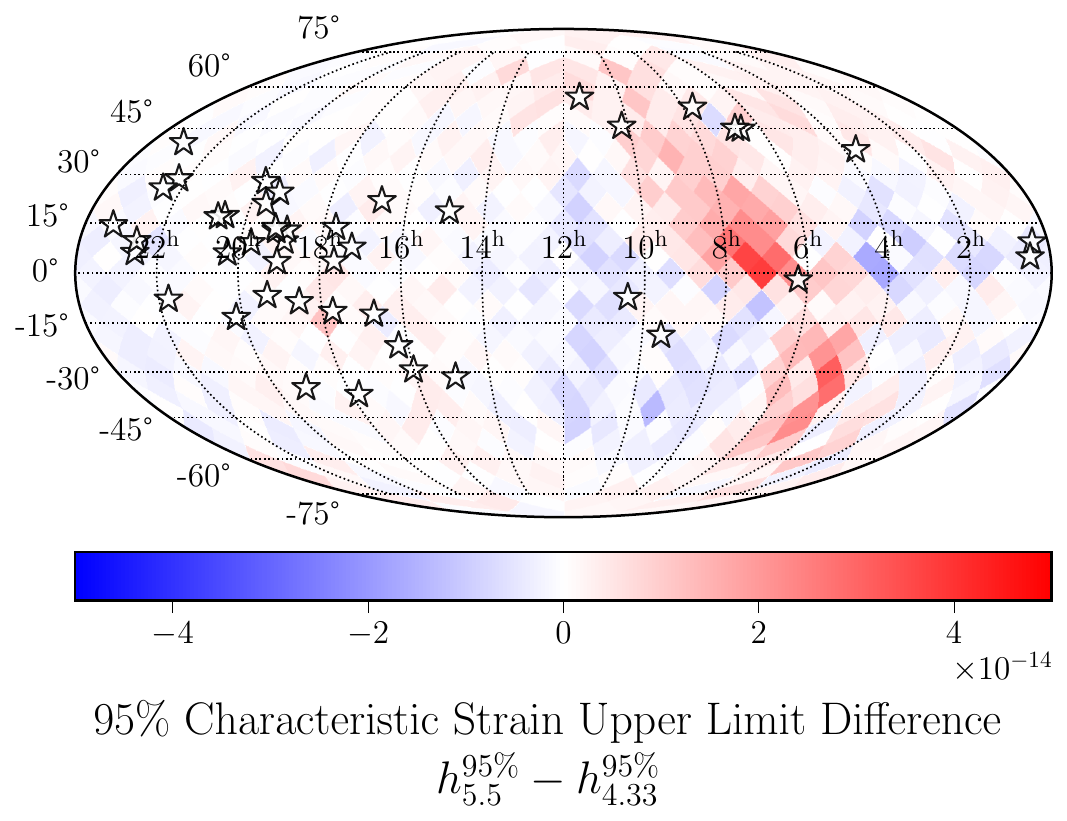}{0.48\textwidth}{}}

    \caption{Top left: The upper limits on memory strain amplitude as a function of skypixel including a CURN law process using a fixed spectral index of $\gamma_{\text{SMBHB}} = 4.33$, as expected for a stochastic gravitational wave background originating from an ensemble of uniformly, isotropically distributed SMBHBs. Top right: The upper limit on memory strain amplitude as a function of sky pixel including a CURN power law process using a fixed spectral index of $\gamma_{\text{MAP}} = 5.5$, the maximum a posteriori value for the detected CURN in \citetalias{arzoumanian_nanograv_2020}. Bottom: The difference between the top two panels. A positive value indicates that the upper limits on strain using a spectral index of $5.5$ are higher. We see that overall, the upper limits change slightly when the red noise model uses the preferred red noise spectral index. However, these changes are localized to a small part of the sky. It can be shown that these upper limit differences can be largely attributed to PSR J0613$-$0200.}
    \label{fig:ULvSky}
    
\end{figure}

\begin{figure}
    \centering
    \includegraphics[width=0.98\textwidth]{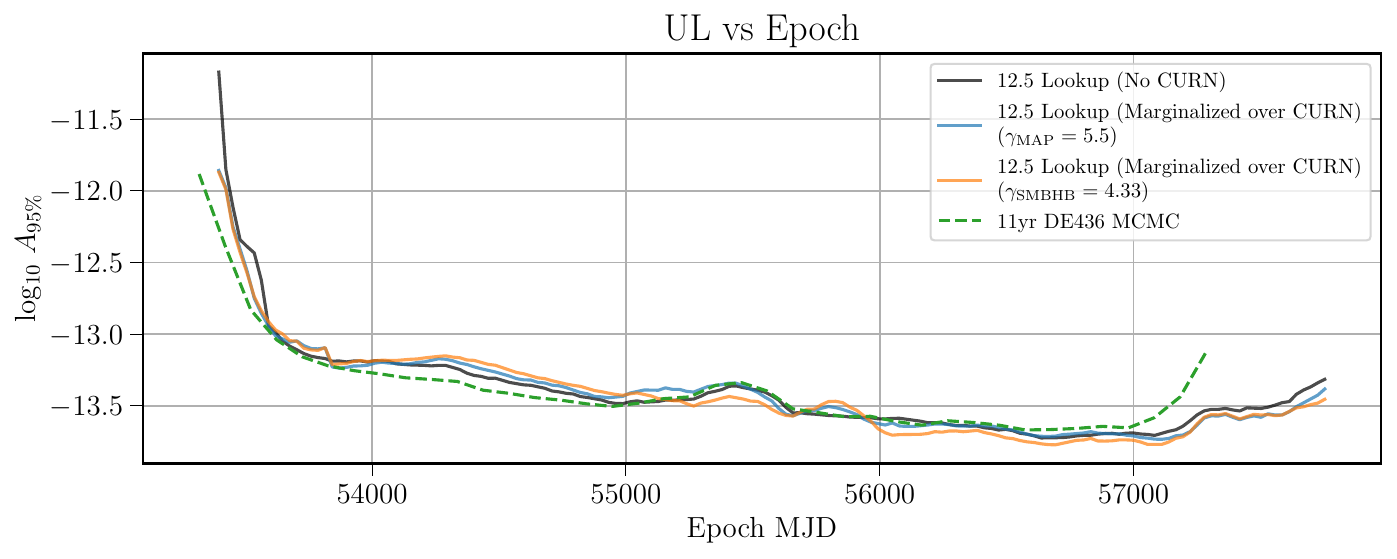}
    \caption{The memory strain amplitude upper limit as a function of burst epoch. The black curve shows the upper limits using a model which does not include CURN. The blue and orange curves show the upper limits using a model which does include a fixed spectral index CURN power law process with $\gamma_{\text{MAP}} = 5.5$ and $\gamma_{\text{SMBHB}} = 4.33$, respectively. We see that the sensitivity of the NANOGrav 12.5-year data set does not give significantly improved upper limits for the first 11 years. However, the additional pulsars and timing data give continuously improving upper limits later in the data set.}
    \label{fig:ULvT}
\end{figure}

\autoref{fig:ULvSky} and \autoref{fig:ULvT} show the upper limits on GW memory strain amplitude in the NANOGrav 12.5-year data set as a function of burst epoch and sky location, respectively.

To compute the upper limits as a function of burst epoch (\autoref{fig:ULvT}), we must compute amplitude posteriors which have uniform priors over the sky and polarization. Thus, we started by splitting up the sky into 48 \texttt{HEALPix}\footnote{https://healpix.sourceforge.io/}\citep{gorski_healpix_2005} sky pixels using \texttt{healpy}\footnote{https://github.com/healpy/healpy} (\texttt{nside=2}) \citep{zonca_healpy_2019}. Then, for each sky pixel, we computed likelihood tables for global GW memory events using \autoref{eqn:factorized_likelihood} and the pulsar-term likelihood tables. Finally, for each trial burst epoch, we took an equal number of samples from the amplitude posteriors from each source-orientation bin at that trial epoch. We then concatenated the samples taken from each of these amplitude posteriors together to form a sky-averaged strain amplitude posterior.

We must sample each source-orientation bin independently to construct our sky-averaged posteriors because of the nature of the memory signal. Our PTA does not have uniformly-distributed pulsars, and as such, there are parts of the sky in which we have little to no sensitivity. If we are not careful about sampling, and instead search over the entire sky simultaneously, our amplitude would be dominated by samples taken from source orientations to which our PTA is completely insensitive. Furthermore, because there is much more prior volume at high amplitudes, these samples would all heavily bias our amplitude posteriors towards higher amplitudes which our PTA has no way of ruling out. This sampling scheme, in which we concatenate samples from different source orientation bins, guarantees that our posterior is marginalized uniformly over the prior \citep{malmquist_relations_1922}. 

\autoref{fig:ULvSky} shows the upper limits on memory strain as a function of sky location and fixed common red-noise spectral index. To obtain these upper limits, we first computed Earth-term lookup tables for 768 \texttt{HEALPix} sky pixels (\texttt{nside = 8}) marginalized over the polarization of the memory wavefront (c.f., the upper limits as a function of burst epoch). For these Earth-term lookup tables, we limit the prior on the burst epoch to the last three years because some pulsars do not have more than three years of data. We can see from this comparison that the upper limits differ slightly depending on the choice of fixed spectral index for the common red noise. While we do not believe these differences are significant, there is a difference pattern that is very similar to the antenna response of a GW memory event around PSR J0613$-$0200. After repeating the same analysis, but omitting this pulsar, the nearby differences largely disappear. This suggests that this pulsar contains a noise feature which is difficult to model accurately using only a red noise power law and white noise. When mismodeled, the excess noise is conflated with a GW memory signal, thus causing the upper limit differences in \autoref{fig:ULvSky}. 


\autoref{fig:ULvT} shows the upper limits on GW memory in the NANOGrav 12.5-year data set plotted as solid curves. We also show the results of the NANOGrav 11-year search for GW memory plotted as a dashed green curve. The upper limits computed from the NANOGrav 12.5-year data set do not improve significantly upon those computed from \citetalias{aggarwal_nanograv_2020} in the overlapping epochs. However, the increased volume of timing data and number of pulsars do clearly result in improvements on the upper limits of approximately half an order of magnitude when compared to early upper limits. Continued observation and growth of this PTA will cause the upper limits in the future to be even lower, and give much more stringent limits on GW memory.

\begin{figure}
    \centering
    \includegraphics[width=\columnwidth]{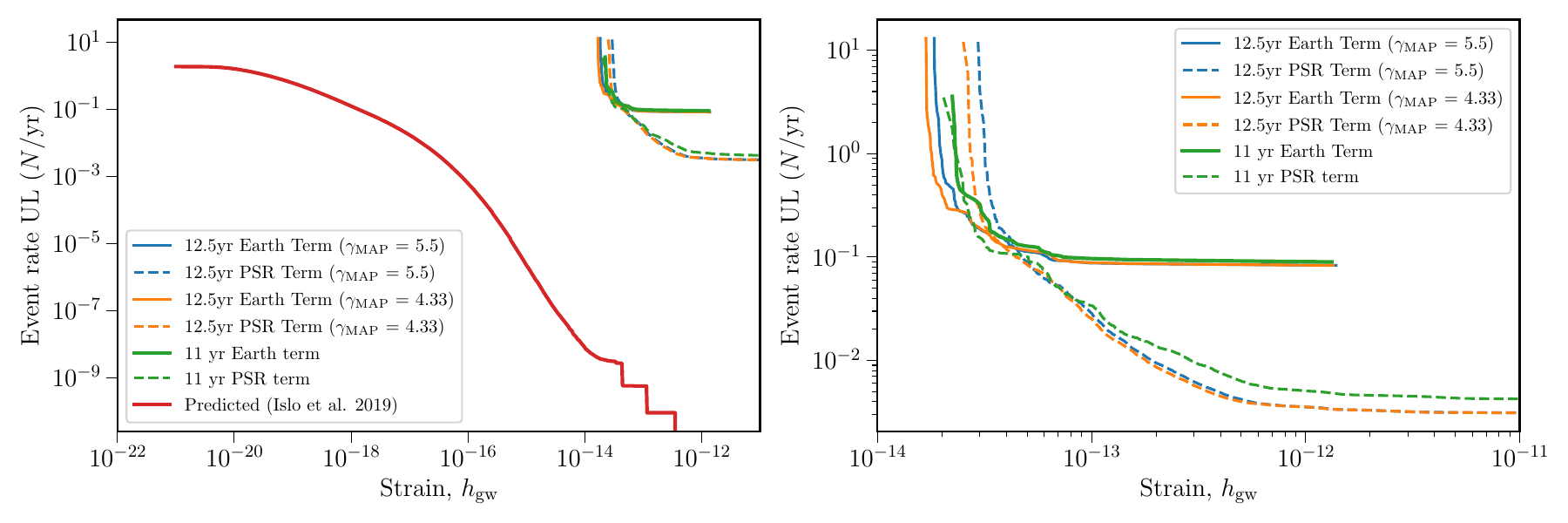}
    \caption{A plot of the predicted rate of SMBHB mergers (red) against the rate upper limits derived from the strain amplitude upper limit as a function of burst epoch. Because the strain upper limits do not improve significantly, the event rate upper limits are correspondingly similar to those presented in \citetalias{aggarwal_nanograv_2020}.}
    \label{fig:UL_rate}
\end{figure}

\autoref{fig:UL_rate} shows the upper limits on the rate of SMBHB mergers that produce GW memory computed using the results shown in \autoref{fig:ULvT}. We do this by counting the number of epochs which have lower strain upper limits than a given fixed strain. From this, we can then constrain the rate of events that have strains at or below this fixed strain. In addition, this figure shows the predicted rate in \citet{islo_prospects_2019}. From the right-hand-side plot, we can see that our rate upper limits do not improve much when compared to the NANOGrav 11-year results. We also include the sky-marginalized pulsar term upper limits. Notably, in this analysis, the Earth term rate upper limits are more constraining than the combined pulsar term upper limits. This indicates that the PTA contains enough pulsars that the sensitivity at low strain amplitudes is no longer dominated by a few pulsars. In addition, the NANOGrav 12.5-year pulsar-term rate upper limits are worse than the 11-year rate upper limits. This is due to the additional red noise model used in the analysis of the 12.5-year data set.

\section{Discussion and Conclusion}


In this paper, we have shown that there is no significant detection of GW memory in the NANOGrav 12.5-year data set. We have therefore set upper limits on the strain amplitude of any GW memory events in the NANOGrav 12.5-year data set in the presence of the CURN detected in \citetalias{arzoumanian_nanograv_2020}. The addition of a CURN to the noise model does not significantly affect the upper limits, but does have some covariance with the GW memory signal. We also see from \autoref{fig:ULvSky} that PSR J0613$-$0200 gives significantly different strain upper limits for sources in its vicinity depending on the choice of spectral index for a CURN process. Furthermore, these differences have a quadrupolar shape, similar to the antenna response of a GW memory signal. This indicates the presence of some excess low-frequency noise in this pulsar which can be conflated with GW memory. From \autoref{fig:ULvT} and \autoref{fig:UL_rate}, we see that the additional data in the 12.5-year data set continue to increase our sensitivity to GW memory, especially later in the data set. This in turn will allow us to continue placing more stringent limits on the rates of memory-producing events.

It is important to remember that the predicted event rate upper limits shown in \autoref{fig:UL_rate} are only for SMBHB mergers. While the prospects for detecting GW memory from SMBHB mergers are low, there are many exotic sources which may be expected to emit GWs and produce GW memory \citep{cutler_gravitational-wave_2014}.  Additionally, pulsar glitches, which are instantaneous changes in the rotational frequencies, produce an almost identical signal as a pulsar-term GW burst with memory \citep{cordes_detecting_2012}. Pulsar glitches have been observed in two millisecond pulsars, PSRs B1821$–$24 and J0613$-$0200 thus far \citep{cognard_microglitch_2004, mckee_glitch_2016}. Of these two, J0613$-$0200 is in NANOGrav's timing data set. However, this glitch occurred before NANOGrav began timing the pulsar, and this glitch should therefore have no effect on this pulsar's timing model or residuals. The pulsar-term upper limits presented in this analysis may be used to set upper limits on glitches in every other pulsar as well. In general, this analysis may be used to cross-validate any detection of any loud GW-producing event.

Finally, the search for GW memory can reveal interesting noise features of a PTA's constituent pulsars. For example, the analysis presented in \citetalias{aggarwal_nanograv_2020} shows that PSRs J1909$-$3744 and J0030+0451 had some excess, unmodeled noise. This analysis shows that there is some excess noise in PSRs J1744$-$1134 and J2043+1711 that conspire to give support for a memory event at MJD 57300. This makes them good candidates for any future studies of noise characteristics, like those presented in \citet{lam_nanograv_2016} and \citet{hazboun_nanograv_2020}. In addition, \autoref{fig:ULvSky} suggests that PSR J0613$-$0200 may have a noise transient which is highly covariant with red noise. Previous work has shown that scattering variations may result in excess correlated noise in pulsar timing data sets \citep{keith_measurement_2013, goncharov_identifying_2021, chalumeau_noise_analysis_2022}. \citet{main_j0613, main_variable_2023} have shown that in particular, data from PSR J0613$-$0200 shows significant evidence of scattering variations. These scattering variations may be the source of the differences in GW memory upper limits in the vicinity of this pulsar when using different CURN spectral indices. As pulsar timing baselines become longer and PTA sensitivity to red noise increases, it will be critically important to explore how strong red noise and components of each pulsar's timing models affect detection prospects of GW memory. 


Overall, the search for GW memory remains a critical part of the GW analysis pipeline because of its use in cross-validation of any potential detections of loud GWs, its ability to reveal unmodeled noise, and the possibility that a GW memory event may reveal exotic GW sources. Continued methods development, as applied to both GW memory and intrinsic pulsar noise,
will allow us to perform more robust searches for these sources using future data sets.

\pagebreak
\emph{Author contributions.}
An alphabetical-order author list was used for this paper in recognition of the fact that a large, decade timescale project such as NANOGrav is necessarily the result of the work of many people. All authors contributed to the activities of the NANOGrav collaboration leading to the work presented here, and reviewed the manuscript, text, and figures prior to the paper's submission.
Additional specific contributions to this paper are as follows. 
ZA, HB, PRB, HTC, MED, PBD, TD, JAE, RDF, ECF, EF, NG-D, PAG, DCG, MLJ, MTL, DRL, RSL, JL, MAM, CN, DJN, TTP, NSP, SMR, KS, IHS, RS, JKS, and SJV developed the NANOGrav 12.5-year data set by performing observations, computing TOAs, checking data, and creating and refining timing models. P.T.B., A.D.J., D.R.M., X.S., and J.P.S. performed the various analyses. J.P.S. and P.T.B. coordinated paper writing.
\par
\emph{Acknowledgments.}
The NANOGrav collaboration receives support from National Science Foundation (NSF) Physics Frontiers Center award numbers 1430284
and 2020265, the Gordon and Betty Moore Foundation, NSF AccelNet award number 2114721, an NSERC Discovery Grant, and CIFAR.
The Arecibo Observatory is a facility of the NSF operated under cooperative agreement (AST-1744119) by the University of Central Florida (UCF) in alliance with Universidad Ana G. M{\'e}ndez (UAGM) and Yang Enterprises (YEI), Inc. The Green Bank Observatory is a facility of the NSF operated under cooperative agreement by Associated Universities, Inc. The National Radio Astronomy Observatory is a facility of the NSF operated under cooperative agreement by Associated Universities, Inc.
L.B. acknowledges support from the National Science Foundation under award AST-1909933 and from the Research Corporation for Science Advancement under Cottrell Scholar Award No. 27553.
P.R.B. is supported by the Science and Technology Facilities Council, grant number ST/W000946/1.
S.B. gratefully acknowledges the support of a Sloan Fellowship, and the support of NSF under award \#1815664.
The work of R.B., R.C., D.D., N.La., X.S., J.P.S., and J.T. is partly supported by the George and Hannah Bolinger Memorial Fund in the College of Science at Oregon State University.
M.C. and S.R.T. acknowledge support from NSF AST-2007993.
M.C. and N.S.P. were supported by the Vanderbilt Initiative in Data Intensive Astrophysics (VIDA) Fellowship.
Support for this work was provided by the NSF through the Grote Reber Fellowship Program administered by Associated Universities, Inc./National Radio Astronomy Observatory.
Support for H.T.C. is provided by NASA through the NASA Hubble Fellowship Program grant \#HST-HF2-51453.001 awarded by the Space Telescope Science Institute, which is operated by the Association of Universities for Research in Astronomy, Inc., for NASA, under contract NAS5-26555.
M.E.D. acknowledges support from the Naval Research Laboratory by NASA under contract S-15633Y.
T.D. and M.T.L. are supported by an NSF Astronomy and Astrophysics Grant (AAG) award number 2009468.
E.C.F. is supported by NASA under award number 80GSFC21M0002.
G.E.F., S.C.S., and S.J.V. are supported by NSF award PHY-2011772.
The Flatiron Institute is supported by the Simons Foundation.
A.D.J. and M.V. acknowledge support from the Caltech and Jet Propulsion Laboratory President's and Director's Research and Development Fund.
A.D.J. acknowledges support from the Sloan Foundation.
N.La. acknowledges the support from Larry W. Martin and Joyce B. O'Neill Endowed Fellowship in the College of Science at Oregon State University.
Part of this research was carried out at the Jet Propulsion Laboratory, California Institute of Technology, under a contract with the National Aeronautics and Space Administration (80NM0018D0004).
D.R.L. and M.A.M. are supported by NSF \#1458952.
M.A.M. is supported by NSF \#2009425.
C.M.F.M. was supported in part by the National Science Foundation under Grants No. NSF PHY-1748958 and AST-2106552.
A.Mi. is supported by the Deutsche Forschungsgemeinschaft under Germany's Excellence Strategy - EXC 2121 Quantum Universe - 390833306.
The Dunlap Institute is funded by an endowment established by the David Dunlap family and the University of Toronto.
K.D.O. was supported in part by NSF Grant No. 2207267.
T.T.P. acknowledges support from the Extragalactic Astrophysics Research Group at Eötvös Loránd University, funded by the Eötvös Loránd Research Network (ELKH), which was used during the development of this research.
S.M.R. and I.H.S. are CIFAR Fellows.
Portions of this work performed at NRL were supported by ONR 6.1 basic research funding.
J.D.R. also acknowledges support from start-up funds from Texas Tech University.
J.S. is supported by an NSF Astronomy and Astrophysics Postdoctoral Fellowship under award AST-2202388, and acknowledges previous support by the NSF under award 1847938.
Pulsar research at UBC is supported by an NSERC Discovery Grant and by CIFAR.
S.R.T. acknowledges support from an NSF CAREER award \#2146016.
C.U. acknowledges support from BGU (Kreitman fellowship), and the Council for Higher Education and Israel Academy of Sciences and Humanities (Excellence fellowship).
C.A.W. acknowledges support from CIERA, the Adler Planetarium, and the Brinson Foundation through a CIERA-Adler postdoctoral fellowship.
O.Y. is supported by the National Science Foundation Graduate Research Fellowship under Grant No. DGE-2139292.

\bibliography{bibliography}
\bibliographystyle{aasjournal}



\end{document}